\documentclass[pra,twocolumn,showpacs]{revtex4}
\usepackage{graphicx}
\def\Journal#1#2#3#4{{#1} {\bf #2}, #3 (#4)}
\def\PR{Phys. Rev.}
\def\PRL{ Phys. Rev. Lett.}
\def\PRA{Phys. Rev. A}
\def\JMP{J. Math. Phys.}

\def\RMP{Rev. Mod. Phys.}

\newcommand{\n}{\nonumber}
\newcommand{\bn}{\begin{eqnarray}}
\newcommand{\en}{\end{eqnarray}}
\newcommand{\h}{\hspace}
\begin{document}
\title {Highly anisotropic Bose-Einstein condensates: crossover to lower dimensionality}
\author{Kunal K. Das}
\affiliation{Optical Sciences Center and Department of Physics,
University of Arizona, Tucson, AZ 85721}\thanks{Present address}
 \affiliation{Department of Physics and
Astronomy, SUNY, Stony Brook, NY 11794-3800}

\date{\today}
\begin{abstract}

We develop a simple analytical model based on a variational method
to explain the properties of trapped cylindrically symmetric
Bose-Einstein condensates (BEC) of varying degrees of anisotropy
well into regimes of effective one dimension (1D) and effective
two dimension (2D). Our results are accurate in regimes where the
Thomas-Fermi approximation breaks down and they are shown to be in
agreement with recent experimental data.

\end{abstract}
 \maketitle

\section{Introduction}

There has been a growing interest in Bose Einstein condensation in
effective lower dimensions, with experimental realization finally
becoming reality \cite{Gorlitz}. Apart from the purely academic
interest, there is a very practical interest as well arising from
the rapidly developing endeavor to create efficient atom
waveguides with potential applications in interferometry and
gyroscopes. For trapped gases effective lower dimensionality means
that excitations along the tightly confined dimension(s) are
energetically not allowed. This is a limiting case of highly
anisotropic condensates which are becoming more common with a new
generation of BEC experiments on surface micro traps
\cite{Ott,Leanhardt}.

Interest in condensates of extreme anisotropy is evident in
numerous current theoretical and experimental work on atomic
waveguides, quasi-2D configurations \cite{Safonov,Anderson} and
the Tonks-Girardeau (TG) 1D limit \cite{Tonks,Girardeau,Lieb} in
which impenetrable bosons show fermionic properties. Magnetic
waveguides for neutral atoms of diverse design have been
constructed in many
laboratories\cite{Leanhardt,Hansch,Bongs,Denschlag,Hinds,Prentiss}.
Bessel beams \cite{Arlt} have been used to produce wave-guide-like
optical confinement. With a slight longitudinal potential these
waveguides can be treated as high aspect ratio traps. Atoms have
been trapped in a 2D optical lattice consisting of quasi-1D
optical wells \cite{Greiner} each with an aspect ratio of up to
about 2000. Work is underway at JILA on various waveguide and
near-waveguide configurations \cite{Kishimoto}. On the theoretical
side the various regimes of quantum degeneracy in both the 1D and
2D limits have been studied in detail in several recent papers
\cite{Olshanii,Kolomeisky,Wright,Petrov1D,Petrov2D,HoMa,Girardeaun,Dunjko,Stoof1,Stoof2}.

But a simple model is lacking which would describe how a
condensate changes as it becomes more anisotropic and eventually
crosses over to effective lower dimensionality.  For condensates
in 3D, of atoms with positive scattering length, a highly
effective and yet essentially simple analytic description was
achieved through the Thomas-Fermi Approximation (TFA) \cite{RMP}.
It neglects the kinetic energy in the Gross-Pitaevskii (GP)
equation which is the mean-field description of the condensate and
it is justified for large condensates with aspect ratios of the
order of unity. But as the aspect ratio deviates farther from
unity, the kinetic energy in the constricted direction becomes
increasingly more important and the TFA does not work so well.
Although there have been numerous, insightful refinements of the
TFA \cite{STFA,LPS,FF,Schuck,Watson} they do not directly improve
on it for highly anisotropic BEC. It is desirable to have a
theoretical model comparable to the TFA in simplicity but one
which is successful in describing condensates from the 3D regime
with increasing degree of anisotropy all the way to regimes of
effective lower dimensionality. That is the objective of this
paper.

For longitudinally homogeneous systems, we have recently
coauthored a study of the crossover from 3D to effective 1D
\cite{crossover}. Here we will consider gases harmonically
confined in all directions and a crossover to effective 2D as
well. We will use a variational approach to obtain analytic
expressions for the chemical potential, total energy, release
energy and also the energy spectrum valid over a wide range of
parameters that include the regime of crossover to lower
dimensions.  In Sec.~\ref{sec:chem} we compare our chemical
potential with accurate numerical solutions of the GP equation and
the Thomas-Fermi expression and then in Sec.~\ref{sec:crossover}
we discuss the crossover regime. In Sec.~\ref{sec:experiment} we
compare the release energy from our model with experimental data
from Ref.~\cite{Gorlitz}.  Then in Sec.~\ref{sec:excited} we
consider the Bogoliubov equations for quasiparticle excitations
and obtain the energy spectrum for geometries close to effective
1D and 2D.

\section{Chemical Potential}\label{sec:chem}

At zero temperature in the lowest order mean-field approximation
the condensate is described by the wave function that minimizes
the GP energy functional \cite{RMP} which in cylindrical
co-ordinates is \bn\label{enfunc} E[\Phi]=N \hbar \omega\int r dr
\int dz \left[\frac{\gamma^{1/3}}{2}(|\nabla_{r}\Phi|^{2}+
r^{2}|\Phi|^{2})\right.\n\\ \left.
+\frac{\gamma^{-2/3}}{2}(|\partial_{z}\Phi|^{2}+z^{2}|\Phi|^{2}) +
   Ng|\Phi|^{4}\right].\en
We have defined the aspect ratio as the ratio of the radial and
the axial trapping frequencies $\gamma=\omega_{r}/\omega_{z}$. The
interaction strength is determined by the scaled scattering length
$g=a/a_{0}$ and by the atom number $N$ which we take to be equal
to the condensate number assuming temperatures close to absolute
zero. The co-ordinates $r$ and $z$ are scaled by the respective
oscillator lengths $r_{0} = \sqrt{\hbar/m\omega_{r}}$ and $z_{0} =
\sqrt{\hbar/m\omega_{z}}$, with the mean oscillator length
$a_{0}=\sqrt{\hbar/m\omega}$ being defined by the geometric mean
frequency, $\omega=(\omega_{r}^{2}\omega_{z})^{1/3}$. The
condensate wavefunction is $(e^{-im\theta}/\sqrt{2\pi})\Phi(r,z)$
with $m=0$ and normalization $\int r dr \int dz
|\Phi(r,z)|^{2}=1$.

Functional minimization $\delta E/\delta\Phi=0$ with the
normalization condition gives the GP equation
\bn \frac{\gamma^{1/3}}{2} \left(-\nabla_{r}^{2} + r^{2} \right) +
\frac{\gamma^{-2/3}}{2} (-\partial_{z}^{2}+z^{2}) + 2Ng
|\Phi|^{2}=\frac{\mu}{\hbar \omega}. \en
 This determines the optimum condensate wavefunction
$\Phi$ and the corresponding chemical potential $\mu$. For
condensates with all its spatial dimensions of comparable
magnitudes the Thomas-Fermi approximation provides a accurate
value of the chemical potential
\bn\label{chemtf} \mu_{TF} = \frac{\hbar \omega}{2}
\left(15Ng\right)^{2/5}.\en
This expression has no dependence on the aspect ratio, whereas we
would expect the chemical potential to change as the aspect ratio
changes. In order to obtain analytic expressions for $\Phi$ and
$\mu$ that will have the correct dependence on the aspect ratio we
take a variational approach with a trial wavefunction containing a
few parameters and then minimize the energy functional  with
respect to them \cite{BaymPethick}. In choosing our trial function
we note that for highly anisotropic traps, in the direction of
weak confinement the condensate size far exceeds the oscillator
length and the kinetic energy (which scales as the inverse square
of the size) becomes negligible, so a Thomas-Fermi form is
appropriate. In the squeezed coordinate the interaction term has
lesser relative importance so that a Gaussian form is suitable.

\begin{figure}\vspace{-5mm}
\includegraphics*[width=\columnwidth,angle=0]{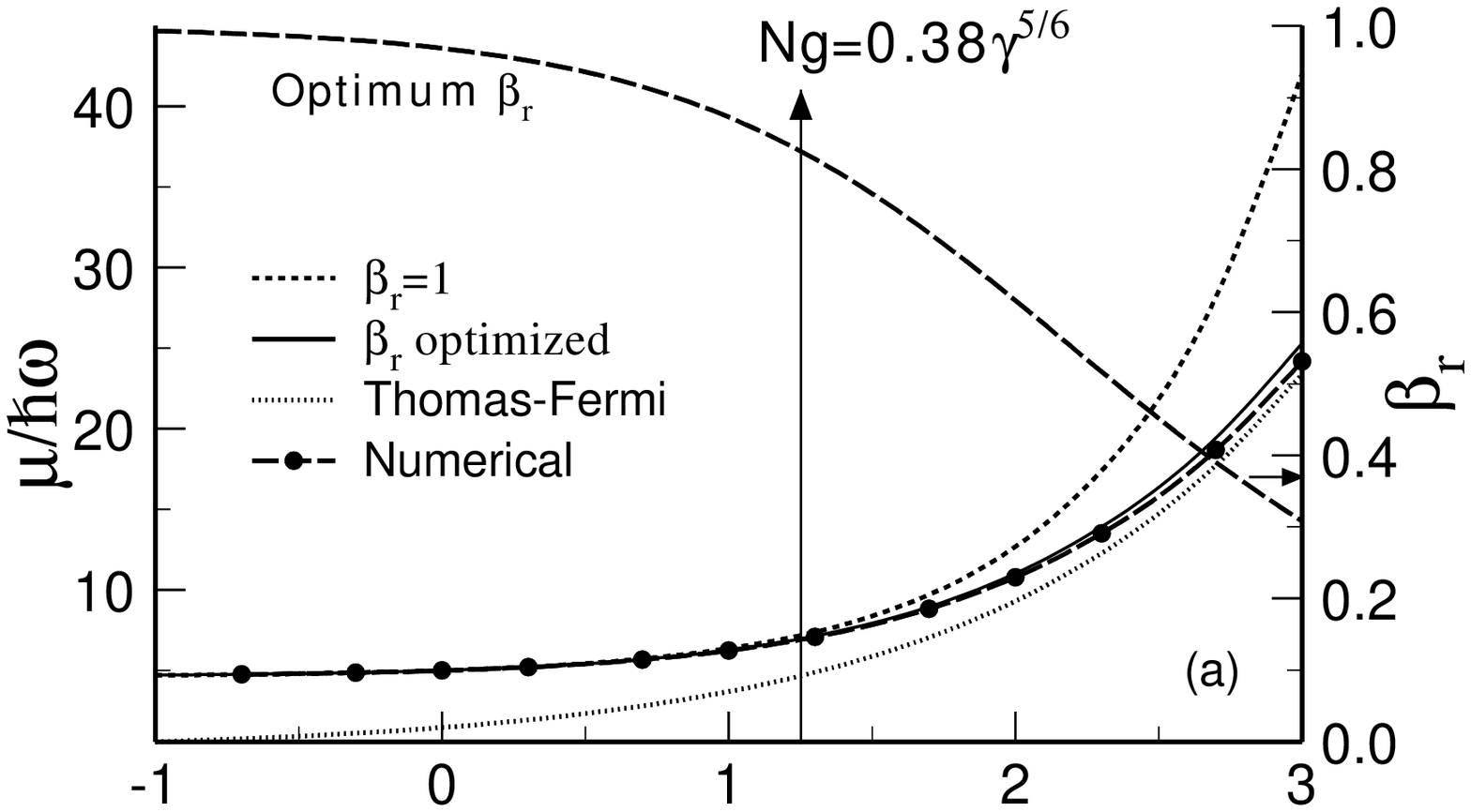}\vspace{-2cm}
\includegraphics*[width=\columnwidth,angle=0]{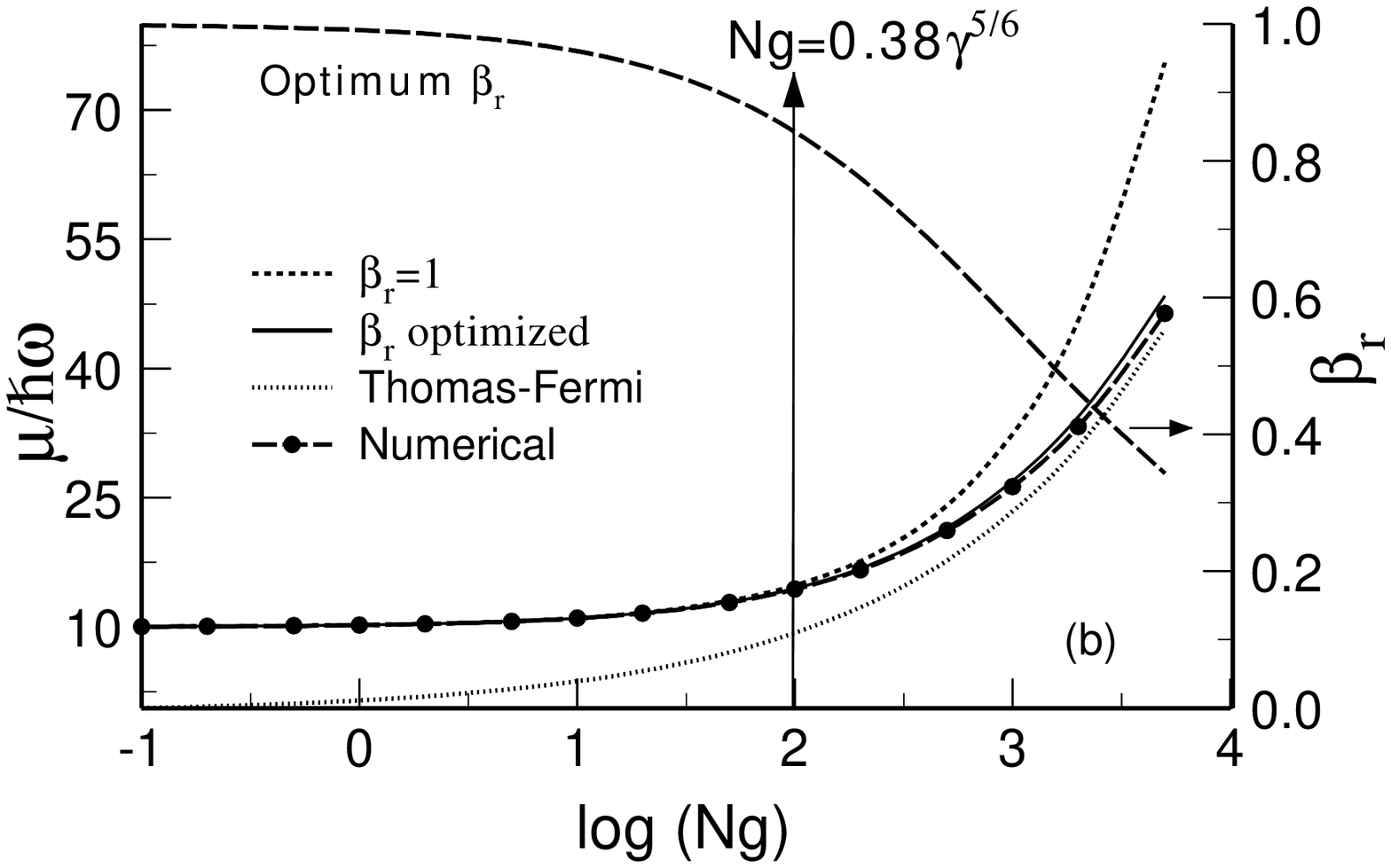}\vspace{-5mm}
\caption{Comparison of chemical potential obtained from various
approaches, for cigar-shaped traps with (a) $\gamma$ = 100
 and (b) $\gamma$ =1000 . The methods are numerical solution
of the Gross-Pitaevskii equation; the Thomas-Fermi approximation;
 analytic expression Eq. (\ref{cigarmu}) with optimized
$\beta_{r}$ and $d$; and the same with $\beta_{r}=1$.  The optimum
value of $\beta_{r}$ is plotted with the right axis scale.}
\label{fig1}
\end{figure}
\subsection{Cigar Geometry}
 For the cigar geometry with $\gamma \gg 1$, we take our
 normalized trial function to be
 \bn\label{cigar}
\Phi_{\rm cigar}(r,z)= \!\left( \frac{3 \beta_{r}}{2 d^{3}}
\right)^{1/2} \h{-3mm} e^{-\beta_{r} r^{2}/2} \sqrt{d^{2}\!-z^{2}}
\ \Theta(d^{2}\! - z^{2}), \en
where $\beta_{r}$ and $d$ are variational parameters and
$\Theta(x)$ is the unit step function. In evaluating the energy
functional, Eq.~(\ref{enfunc}), for this trial function, we
recognize that the kinetic energy in the transverse direction
should be included, but not that in the axial direction. We thus
obtain \bn \label{pbeta}\frac{E[d,\beta_{r}]}{N\hbar \omega}=
\left[ \frac{\gamma^{1/3}}{2}\left(\beta_{r} + \frac{1}{\beta_{r}}
\right) +\frac{\gamma^{-2/3} d^{2}}{10} + \frac{3\beta_{r} Ng}{5d}
\right].\en Minimizing $E$ with respect to $d$ and $\beta_{r}$
gives
\begin{eqnarray}\label{dbetarho}
d^{3} =3 \beta_{r} Ng \gamma^{2/3}; \h{1cm}
\frac{1}{\beta_{r}^{2}}=1 + \frac{6Ng}{5d} \gamma^{-1/3}.
\end{eqnarray}
These coupled equations can be easily solved numerically, and the
chemical potential corresponding to the optimum parameters is
 \bn \label{cigarmu}\mu = \hbar \omega \left[
\frac{\gamma^{1/3}}{2} \left( \beta_{r} + \frac{1}{\beta_{r}}
\right) + \frac{1}{2} \left(\frac{3\beta_{r}
Ng}{\gamma^{1/3}}\right)^{2/3}
\right] \h{4mm}\n \\
= \frac{\hbar \omega_{r}}{2} \left( \beta_{r} +
\frac{1}{\beta_{r}} \right) + \frac{1}{2} \left( 3 \beta_{r} N a
\omega_{r} \omega_{z} \hbar\sqrt{m} \right)^{2/3}. \en
This satisfies the thermodynamic relation $\mu = \partial
E/\partial N$ as expected. It is apparent from the unscaled form
that the last term is $\propto \omega_{z} ^{2/3}$ and hence
vanishes in the limit $\omega_{z} \rightarrow 0$, also in that
limit $\beta_{r}\rightarrow 1$ so that the chemical potential
$\mu$ equals the transverse ground-state energy $\hbar\omega_{r}$
as we would expect.  In fact for very elongated traps, to a very
good approximation we can set $\beta_{r}= 1$ and  Eq.
(\ref{cigarmu}) reduces to a self-contained \emph{analytic
expressions for $\mu$ dependent only on N,a and $\gamma$} all of
which are measurable parameters of the system.

In order to test the accuracy and validity of our approximation we
compare the analytic expressions for the chemical potential with
accurate numerical solutions of the GP equation, obtained using a
discrete variable representation (DVR) mesh in $r$ (Laguerre DVR)
and $z$ (Hermite DVR) \cite{BayeHeenan,SF,BFBS}. We checked for
convergence of the numerical results as a function of mesh size
and range. Typically, 1,000 to 1,500 mesh points sufficed.

In Fig.~\ref{fig1} we plot the chemical potential computed in
various ways for cigar traps of two different aspect ratios. It is
clear that the variational chemical potential using expression
(\ref{cigarmu}) with optimized $\beta_{r}$ closely follows the
numerically computed chemical potential over a large range of
 $Ng$.  For low linear
densities, the expression with $\beta_{r} = 1$ is sufficient but
it fails as the density increases. On the other hand the
Thomas-Fermi expression (\ref{chemtf}) is accurate at high
densities and breaks down at low densities since it approaches
zero while the correct $\mu$ should approach the zero-point
energy.

\subsection{Pancake Geometry}
\begin{figure}\vspace{-5mm}
\includegraphics*[width=\columnwidth,angle=0]{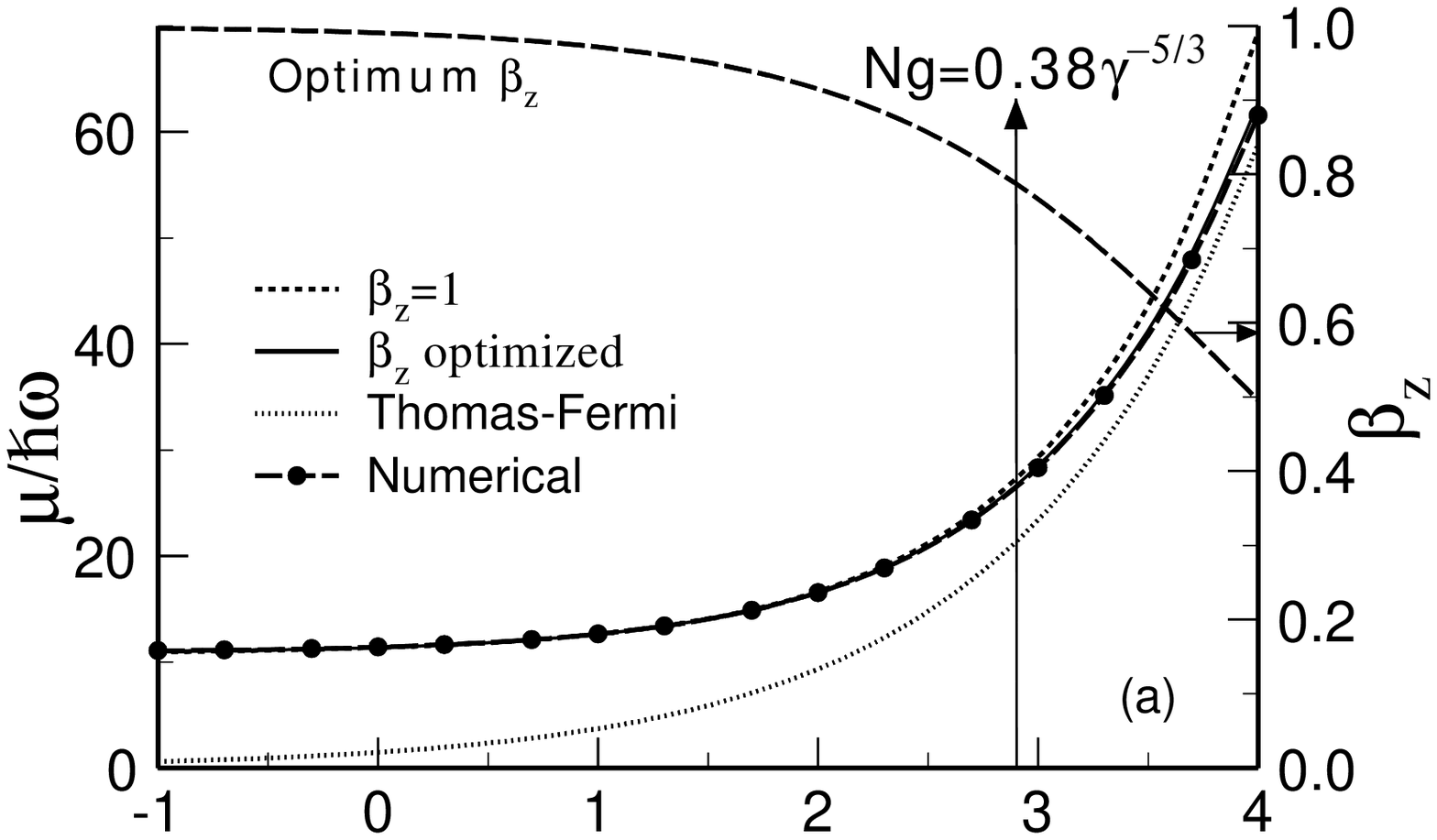}\vspace{-2cm}
\includegraphics*[width=\columnwidth,angle=0]{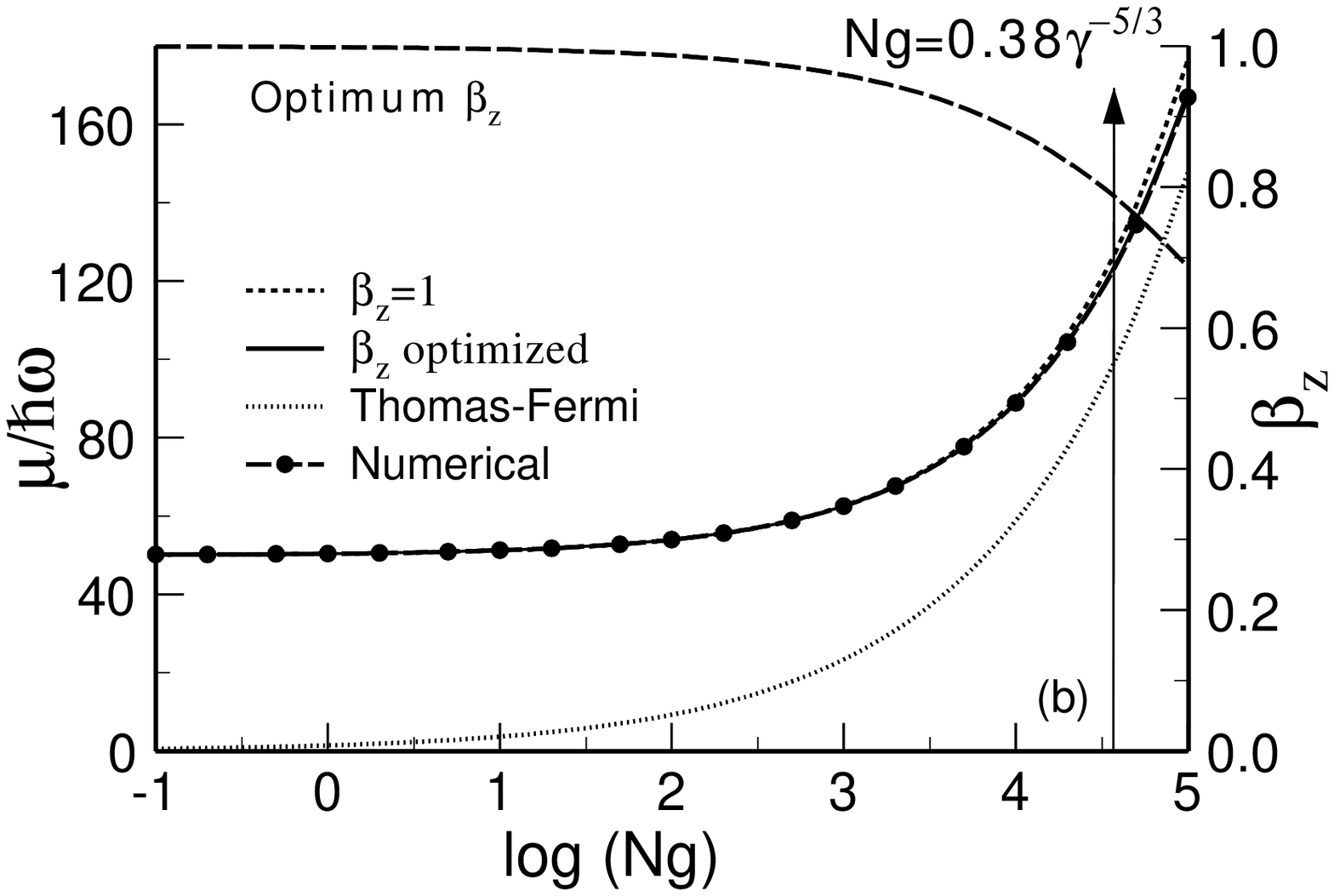}\vspace{-5mm}
\caption{The chemical potential for pancake shaped traps obtained
from various approaches, for aspect ratios (a) $\gamma$ = 0.01
(top) and (b)$\gamma$ = 0.001. In both cases, the result with
optimized $b$ and $\beta_{z}$ is essentially indistinguishable
from the numerical solution of the GP equation.} \label{fig2}
\end{figure}

 On carrying  through a similar analysis for the pancake
geometry with $\gamma\ll 1$, taking as trial function \bn
\Phi_{\rm pan} = \frac{2}{b^{2}} \left( \frac{\beta_{z}}{\pi}
\right)^{1/4} (b^{2} - r^{2})^{1/2} e^{-\beta_{z} z^{2}/2}\
\Theta(b^{2}\! - r^{2}),\en
we find that the energy functional, neglecting the transverse
kinetic energy, is
 \bn \label{efunc:pan}\frac{E[b,\beta_{z}]}{N\hbar
\omega}=\left[\frac{\gamma^{1/3}b^2}{6}+ \frac{\gamma^{-2/3}}{4}
\left(\beta_{z}+\frac{1}{\beta_{z}}\right)+ \frac{8Ng }{3b^2}
\sqrt{\frac{\beta_{z}}{2\pi}}\right].\en
The equations for the optimum parameters that minimize the energy
are
 \bn b^{4} =
16\frac{Ng}{\gamma^{1/3}}\sqrt{\frac{\beta_{z}}{2\pi}},\h{1cm}
\frac{1}{\beta_{z}^{2}}=1+ \frac{16}{3b^2}
\frac{Ng\gamma^{2/3}}{\sqrt{2\pi\beta_{z}}}, \en
 and the corresponding expression for the chemical potential
\bn \label{panmu}\mu=\hbar \omega \left[ \frac{\gamma^{-2/3}}{4}
\left(\beta_{z}+ \frac{1}{\beta_{z}} \right) +
\left(\frac{8\beta_{z}}{\pi}\right)^{1/4}\h{-4mm}\sqrt{Ng\gamma^{1/3}}
\right].\en
As we did for the cigar geometry, we plot the chemical potential
evaluated in various ways in Fig.~\ref{fig2}. The deviation of the
Thomas-Fermi expression from the correct expression for $\mu$ is
more pronounced in this geometry while our variational expression
from Eq.~(\ref{panmu}) reproduces almost exactly the numerical
solution of the GP equation over the range of $Ng$ shown. Even the
analytic expression for $\mu$ with $\beta_{z}=1$ is accurate to
large values of $Ng$. The reason for the better agreement is
larger relative importance of the kinetic energy in the tightly
confined direction. Note also that in both Fig.~\ref{fig1} and
Fig.~\ref{fig2}, differences between the TFA and the numerical
results persist to larger values of $Ng$ as the anisotropy
increases i.e. $\gamma$ deviates more from unity.

\begin{figure}
\includegraphics*[width=\columnwidth,angle=0]{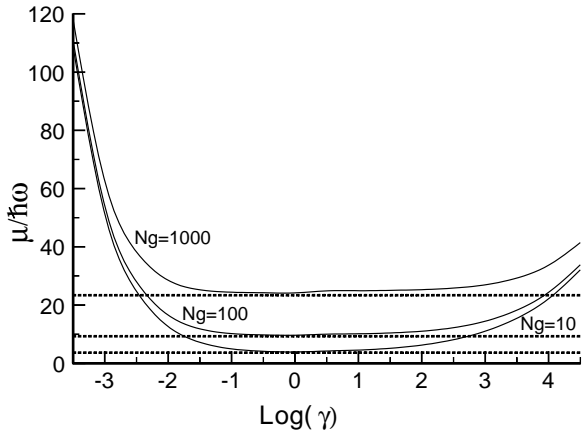}\vspace{-5mm}
\caption{The variationally optimized chemical potential $\mu$
 is plotted as a function of the
aspect ratio $\gamma$ for different mean field strengths. For
$\gamma>1$ Eq.~(\ref{cigarmu}) is used and for $\gamma<1$
Eq.~(\ref{panmu}). The horizontal dashed lines, that represent the
Thomas-Fermi chemical potential $\mu_{TF}$, are tangential to the
corresponding variational $\mu$ at their lowest points at
$\gamma=1$. } \label{fig3}
\end{figure}

In Fig.~\ref{fig3} we plot the chemical potential obtained from
our variational expressions in Eqs.~(\ref{cigarmu}) and
(\ref{panmu}) as a function of the aspect ratio $\gamma$, over
several orders of magnitudes that vary from very elongated cigar
condensates to extremely oblate pancake shapes. Strong variation
with the aspect ratio is apparent. In the same figure the
Thomas-Fermi chemical potential $\mu_{TF}$ (\ref{chemtf}) appears
as flat horizontal lines tangential at $\gamma=1$ to the lines
corresponding to our calculated $\mu$. Thus our expressions for
the chemical potential agree with the Thomas-Fermi value for
spherical symmetry.

\section{Crossover to lower dimensions}\label{sec:crossover}

The variational parameters we have used to obtain the chemical
potential also provide a measure of the effective dimensionality
of the condensate. A trapped Bose gas is considered to be in
effective lower dimension if excitations in the tightly confined
dimension are frozen.  Thus a criteria for crossover to lower
dimensionality \cite{Gorlitz} is when the interaction energy per
particle is comparable to the energy ($\sim \hbar\omega_{tight}$)
to excite the first excited mode in the tightly confined
direction. For a weakly interacting Bose gas at low densities the
interaction energy is roughly equal to the Thomas-Fermi or 3D
chemical potential so that the condition for lower dimensionality
is $\mu_{TF}<\hbar\omega_{tight}$ which in our units become
\bn gN_{1D} < \sqrt{\frac{32}{225}}\ \gamma^{5/6}\simeq
0.38\times\gamma^{5/6}\h{5mm}\n\\
 gN_{2D}<\sqrt{\frac{32}{225}}\ \gamma^{-5/3}\simeq 0.38\times\gamma^{-5/3}.
 \label{crosscrit}\en
In  Figs.~\ref{fig1} and \ref{fig2}, we have plotted the optimum
value of $\beta_{r}$ and $\beta_{z}$ respectively along the right
axis. The values corresponding to the crossover to one dimension
$gN_{1D} = 0.38\gamma^{5/6}$ and to two dimension
$gN_{2D}<0.38\gamma^{-5/3}$ are indicated.  We see that this
roughly corresponds to $\beta_{r}\sim 0.8$ and $\beta_{z}\sim
0.8$.  Thus the values of $\beta_{r}$ and $\beta_{z}$ give a
measure of the dimensionality of the system: when $\beta_{r}\sim
1$ the system is effectively one-dimensional since the transverse
profile of the condensate coincides with that of the transverse
ground state, likewise when $\beta_{z}\sim 1$ the system is
effectively two dimensional.  As these parameters deviate from
unity the system approaches 3D.  It is worth noting at this point
that in Ref.~\cite{crossover} a similar criterion was used to
study the 1D-3D crossover for a homogeneous system, although the
parameter used was different, being the effective 1D interaction
strength.

The 1D case is of particular interest because at extreme low
density and tight confinement the gas becomes an impenetrable
Tonks-Girardeau gas \cite{Tonks,Girardeau}.  That regime is a
subset of the regime of one-dimensionality. A wavefunction for a
gas of such impenetrable bosons can be mapped to that of a gas of
free fermions.

With slight rearrangement the criterion (\ref{crosscrit}) we have
used above for one dimensionality can be written as $N <
0.38\times (r_{0}/a)\gamma$.  The impenetrable regime requires a
much more stringent condition $k_{F}r_{0}^{2}/a\ll 1$
\cite{Olshanii}, where $k_F$ is the Fermi wavevector of the
equivalent Fermi system.  For a large enough number of atoms and
at zero temperature, the asymptotic form of the harmonic
oscillator modes provides a estimate for the Fermi wavevector
$k_{F}=\sqrt{2N}/z_{0}$, so that the condition for an impenetrable
gas translates to $N\ll \frac{1}{2}\gamma (a/r_{0})^{2}$.  For
typical traps $a\ll r_{0}$ which means that the particle number
needs to be much smaller than that required for
one-dimensionality. Thus it follows from our considerations above
that the transverse profile in the impenetrable regime has to be
that of the ground state of the transverse trap potential.

Our variational functions however cannot be applied to the
Tonks-Girardeau regime, since the Thomas-Fermi profile in the
axial direction has to be replaced by a square-root of a parabola
\cite{Kolomeisky}, and the axial energy is simply the Fermi energy
for N particles in 1D.  Thus while the energies and chemical
potential we have derived can be applied over a wide range of
parameters spanning effective 1D and 2D, they are not meant to
describe the impenetrable regime.

\section{Comparison with Experiment}\label{sec:experiment}

In the experiment in Ref.~\cite{Gorlitz} where BEC in effective
lower dimensions was achieved, the crossover from a 3D to lower
dimensionality was deduced (i) by observing a sudden change in the
aspect ratio of the released condensate when the number of atoms
was lowered below a certain value, and (ii) by observing a
saturation of the release energy at the zero-point kinetic energy
in the tightly confined direction. Since the measurement of the
aspect ratio was done several milliseconds after release from the
trap, a proper theoretical description would involve the dynamics
\cite{CastinDum} of the expansion of the condensate which we do
not consider in this paper.  However due to conservation of energy
we can estimate the release energy in our model and compare it
with the experimentally observed values.

The release energy is the energy of the system after the traps are
switched off, which is just the sum of the kinetic and the
interaction energies of the condensate before release. For the
cigar geometry $\gamma \gg 1$, the variational calculation in
Sec.~\ref{sec:chem} gives the release energy per particle
\bn\label{releasec} E_{\rm rel} = \frac{\hbar\omega_{r}\beta_{r}
}{2} + \frac{\hbar\omega_{z}d^{2}}{5}, \en
where we use the optimized parameters $\beta_{r}$ and $d$ from
Eq.~(\ref{dbetarho}).  The release energy in \cite{Gorlitz} is
plotted as a function of the half-length (Z) of the condensate. It
turns out that the axial expansion in the 1D experiment  was
negligible within the time of flight till measurement so that for
comparison we simply need the initial half-length before release.

 Considering the form of our trial
function (\ref{cigar}) it is tempting to identify $d$ as the
half-length of the condensate. However, this would not be correct
since the trial function assumes a \emph{uniform} cylindrical
cross-section along the length of the condensate, whereas as we
approach the 3D regime, the condensate tends to be an ellipsoid,
so that the center will tend to bulge out more than the ends.

 One could invoke arguments based on geometry to extract the half length from our
expression for $d$. However a much easier way to do that is to
make sure that our expression for the half-length has the correct
form in the 1D limit and the 3D (Thomas-Fermi) limit
\bn Z_{1D}=z_{0}\left[3Ng\gamma^{2/3}\right]^{1/3}\h{-1mm},\h{3mm}
Z_{3D}=z_{0}\left[15Ng\gamma^{5/3}\right]^{1/5}.\en
The 1D limit is obtained trivially $d(\beta_{r}=1)\rightarrow
Z_{1D}$. In considering the 3D limit we note that
$1/\sqrt{\beta_r}$ is the width of transverse variational profile,
and the second term in the expression (\ref{dbetarho}) for
$\beta_r$ measures the deviation from 1D; as the condensate moves
away from the 1D regime and approaches a ellipsoidal shape, that
deviation would be maximum at the center and negligible at the
end. Thus we take an average value for the deviation and define
the modified parameter $\bar{\beta}_r$
\bn \frac{1}{\bar{\beta}_{r}^{2}}=1+\frac{3Ng}{5d}
\gamma^{-1/3}\en
It is easy to see that in the 3D limit  $\bar{\beta}_{r}\ll 1$, we
have $\bar{\beta}^{2}\simeq 5d\gamma^{1/3}/3Ng$ and we get the
correct limit $d(\bar{\beta}_r)\rightarrow Z_{3D}$, while we get
the correct 1D limit as well $d(\bar{\beta}_r=1)\rightarrow
Z_{1D}$. Thus in the intermediate regime we expect the half-length
to be given to a good approximation by
 \bn\label{hl} Z =3 Ng\bar{\beta}_r\gamma^{2/3}.\en

\begin{figure}
\includegraphics*[width=\columnwidth,angle=0]{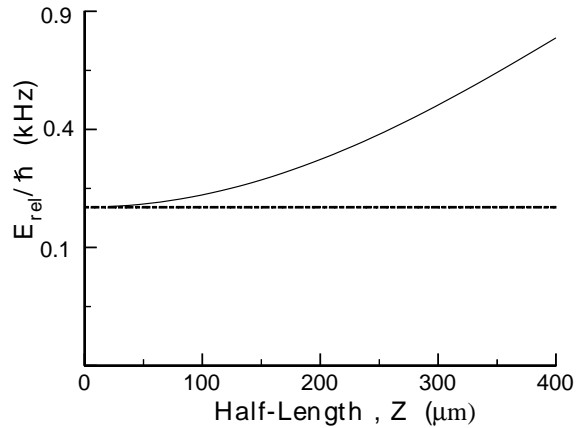}\vspace{-5mm}
\caption{The release energy per particle $E_{rel}/\hbar$ plotted
as a function of the half-length of a cigar-shaped condensate of
trapping frequencies $\omega_{r}=2\pi\times 360$ Hz and
$\omega_{z}=2\pi\times 3.5$ Hz. Compare with Fig.~3(b) in
Ref.~\cite{Gorlitz}. The horizonal line represents the transverse
zero-point energy.} \label{fig4}
\end{figure}

We stress that the release energy should be evaluated with the
optimized variational parameters from Eq.~(\ref{dbetarho}). The
variational process optimizes energy and not condensate dimensions
and hence there is no inconsistency.  In Fig.~\ref{fig4} we plot
the release energy per particle from Eq.~(\ref{releasec}) versus
the half length from Eq.~(\ref{hl}) for the parameters
corresponding to the 1D experiment in Ref.~\cite{Gorlitz}: sodium
atoms in a magnetic trap with radial frequency
$\omega_{r}=2\pi\times 360$ Hz and axial frequency
$\omega_{z}=2\pi\times 3.5$ Hz, so that the aspect ratio is
$\gamma\simeq 103$. A comparison with plot (3b) presented in
Ref.~\cite{Gorlitz} shows that our expression for the release
energy closely follows their measured data for the release energy;
as the system approaches effective 1D the saturation of the
release energy at the radial zero-point energy is clear.

In the pancake geometry the release energy per particle from our
variational calculations is
\bn\label{releasep} E_{\rm rel} = \frac{\hbar\omega_{z}\beta_{r}
}{4} + \frac{\hbar\omega_{r}b^{2}}{6} \en
Considering the agreement in the cigar case we expect this
expression to agree well with experimentally measured release
energy in this geometry. We will not do an actual comparison here
for two reasons: First the above expression is for strictly
cylindrical geometry whereas the trap configuration in the 2D
experiment in \cite{Gorlitz} was not. Secondly unlike the 1D
experiment there is significant expansion of the condensate in all
directions till the time of measurement which involves the
dynamics of expansion which is not a topic of this paper.

\section{Excited States}\label{sec:excited}
In Sec.~\ref{sec:chem} we found the chemical potential to be
increasingly different from the Thomas-Fermi value as the
anisotropy of the system increases; since the quasiparticle
spectrum depends on the chemical potential we should expect some
consequent changes. The energy spectra for purely 1D and 2D system
have been known for sometime \cite{HoMa},  also solutions have
been found for cylindrical geometry in 3D using TFA in all
directions \cite{Ohberg,Fliesser,Stringari2}.  We will consider
here how the spectrum changes as the system deviates from
effective 1D or 2D.

Consider the Bogoliubov equations for the quasiparticle
excitations; as with the ground state we assume factorized spatial
dependence of the quasi-particle modes $U(r,z)$ and $V(r,z)$. We
first discuss the pancake geometry. The strong axial confinement
makes it reasonable to assume that $U,V$ have the same
$z$-dependence as the condensate:
 \bn
U_{nm}(r,z)\!=\!R_{nm}^{+}(r)\zeta(z),\h{2mm}
V_{nm}(r,z)\!=\!R_{nm}^{-}(r)\zeta(z), \n\\
\Phi(r,z)=\phi(r)\zeta(z)\h{3mm} {\rm with}\h{2mm}\zeta(z)=\left(
\frac{\beta_{z}}{\pi} \right)^{1/4} \h{-2mm}e^{-\beta_{z}
z^{2}/2}.\en This allows us to integrate out the axial dependence.
On using our expression Eq. (\ref{panmu}) for the chemical
potential the terms reflecting axial kinetic and potential energy
drop out both in the GP and the Bogoliubov equations and they take
the form
 \bn\label{bog:p3} \left[-\nabla_{r}^{2}-(b^{2}-r^{2})
+4g_{2D}|\phi|^{2}\right]\phi=0 \h{1.7cm}\n\\
\left[-\nabla_{r}^{2}-(b^{2}-r^{2})+\frac{m^2}{r^2}+
8g_{2D}|\phi|^{2} \right]R_{nm}^{\pm} \h{1cm}\n\\+
4g_{2D}|\phi|^{2}R_{nm}^{\mp} =\pm2\,\epsilon_{nm}\,R_{nm}^{\pm},
\en Here $g_{2D}=b^{4}/16$ defines an effective transverse
two-body interaction strength and $\hbar\omega_{r}\epsilon_{nm}$
are the transverse normal mode energies.

 The sum and
difference of the Bogoliubov equations can be converted to normal
mode equations for the transverse density and phase fluctuations
\bn \left[-\frac{\nabla_{r}}{|\phi|^{2}}\left[
|\phi|^{2}\nabla_{r}\right] +\frac{m^2}{r^2}+ 4g_{2D}|\phi|^{2}
\right]\frac{\rho_{nm}}{\rho_{0}}
=4i\,\epsilon_{nm}\theta_{n}\n\\
\left[-\frac{\nabla_{r}}{|\phi|^{2}}\left[
|\phi|^{2}\nabla_{r}\right] +\frac{m^2}{r^2}\right]\theta_{nm}
=-i\epsilon_{nm}\frac{\rho_{nm}}{\rho_{0}}. \en
 This is achieved by writing the small amplitude fluctuations
of the condensate wavefunction as
$\delta\phi=\sqrt{\rho}e^{i\delta\theta}-\sqrt{\rho_{0}}$ about
its equilibrium $\phi_{0}=\sqrt{\rho_{0}}$, whereby we get the
well known forms for the linearized density and phase fluctuations
\cite{Fetter}
 \bn \delta\rho=\sqrt{\rho_{0}}(\delta\phi+\delta\phi^{*})\h{1cm}
\delta\theta= \frac{-i}{2\sqrt{\rho_{0}}}(\delta\phi
-\delta\phi^{*})\en
 and the corresponding normal mode amplitudes
 \bn \rho_{nm}\!=\!\sqrt{\rho_{0}}(R^{+}_{nm}\!\!-\!R^{-}_{nm}),\h{2mm} \theta_{nm}
\!=\!\frac{-i}{2\sqrt{\rho_{0}}}(R^{+}_{nm}\!\!-\!R^{-}_{nm}).
 \en
 Separating the equations and neglecting higher than second order
derivatives, being small due to the large transverse size, we find
that the phase fluctuations obey the same hydrodynamic equation as
density fluctuations   \bn \left[-\nabla_{r}\left[
g_{2D}\rho_{0}\nabla_{r}\right]
+g_{2D}\rho_{0}\frac{m^2}{r^2}\right]\frac{W_{nm}}{r^{|m|}}
=\epsilon_{nm}^{2}\frac{W_{nm}}{r^{|m|}} \en
 where $\rho_{nm}(r)=\theta_{nm}(r)=r^{|m|}W_{nm}(r)$.
Taking the Thomas-Fermi expression for the transverse condensate
  density this can be reduced to a hypergeometric equation by defining
  $x=r^{2}/b^{2}$:
 \bn x(1-x)W_{nm}''+\left[(|m|+1)-(|m|+2)x\right]W_{nm}'
\n\\+\frac{\epsilon_{nm}^{2}-|m|}{2}W_{nm}=0\en
 which has the eigenvalues and corresponding  eigenstates (unnormalized)
\bn\label{2denergy}
\epsilon_{nm}\!=\hbar\omega_{r}\sqrt{2n(n\!+\!m\!+\!1)+m}
\h{4mm}n,m=0,1,2,\cdots,\n\\
R_{nm}(r)\!=\!r^{m}\!\sum_{l=0}^{n}\frac{(n+m+l)!}{(n+m)!(m+l)!}\left(
\begin{array}{c}n \\ l \end{array}
\right)\!\!\left(-\frac{r^{2}}{b^{2}}\right)^{l}\h{2mm}\en

For a similar analysis of the cigar geometry, we define
 \bn
U_{n}(r,z)=\psi(r)Z_{n}^{+}(z),\h{5mm}
V_{n}(r,z)=\psi(r)Z_{n}^{-}(z), \h{2mm}\n\\
\Phi(r,z)=\psi(r)\varphi(z)\h{5mm} {\rm
with}\h{2mm}\psi(r)=\sqrt{2\beta_{r}}\,e^{-\beta_{r}r^{2}/2}\en
 and in the limit of extended axial dimension  we find that
the transverse density and phase fluctuations both satisfy the
hydrodynamic equation
 \bn \label{bog:c1} \left[-\partial_{z}(
g_{1D}\rho_{0}\partial_{z})\right]P_{n}(z/d)=\epsilon_{n}^{2}
P_{n}(z/d) \en
with $\rho_{n}(z)=\theta_{n}(z)=P_{n}(z/d)$ and an effective 1D
interaction strength $g_{1D}=d^{3}/3$. For the Thomas-Fermi
condensate density in the longitudinal direction, $P_{n}(z/d)$ are
Legendre polynomials with eigenvalues
 \bn\label{1denergy}
\epsilon_{n}=\hbar\omega_{z}\sqrt{\frac{1}{2}n(n+1)}\h{5mm}n=0,1,2\cdots.\en

The expressions for the normal mode energies and eigenstates in
(\ref{2denergy}) and (\ref{1denergy}) are identical with results
for effective 1D and 2D obtained in previous works
\cite{Petrov1D,Petrov2D,HoMa}. However the total energy for each
mode has to include the energy due to the tightly confined
direction, which in effective 2D is simply $\hbar\omega_{z}/2$ and
in 1D is $\hbar\omega_{r}/2$. However as the system deviates from
lower dimensionality we see that contribution will change and the
total energy will be
\bn\label{totalen}
2D:E_{nm}\!=\frac{\hbar\omega_{z}}{4}\left(\beta_{z}+\frac{1}{\beta_{z}}\right)+
\hbar\omega_{r}\sqrt{2n(n\!+\!m\!+\!1)+m}\n\\
1D:E_{n}=\frac{\hbar\omega_{r}}{2}\left(\beta_{r}+\frac{1}{\beta_{r}}\right)
+\hbar\omega_{z}\sqrt{\frac{1}{2}n(n+1)}\h{1.3cm} \en
with $n,m=0,1,2,\cdots$.  These expressions for the energy
spectrum would apply when the system has deviated form effective
lower dimensionality but has not completely attained Thomas-Fermi
behavior in all directions. In this context we may point out that
for a 3D cylindrically symmetric system well described by
Thomas-Fermi in all dimensions, the energy spectrum is quite
different from the 1D spectrum \cite{Stringari2,Fliesser}.

Moreover the behavior of our eigenfunctions in the weakly confined
dimension(s) depend on the behavior in the tightly confined
dimension(s) through the interdependence of the variational
parameters.  Thus in the crossover regime where the $\beta_{r}$
and $\beta_{z}$ start to deviate significantly from unity, the
shapes of the eigenmodes even in the dimension(s) of weak
confinement will begin to change from what they are in effective
1D or 2D.

\section{Conclusion}

We  have developed a simple model based on a variational approach
which can accurately describe the properties of a Bose Einstein
condensate from the 3D Thomas-Fermi regime through various degrees
of anisotropy well into the regimes of effective 1D and 2D.  The
advantage of the model lies in its simplicity, so that analytic
expressions could be derived for several physical quantities, and
that it can explain some of the results of experiments on such
systems without having to resort to complicated numerical
computations.

In particular we have found expressions for the chemical potential
which are valid for cylindrical condensates for all degrees of
anisotropy even where the Thomas Fermi expression is completely
inadequate.  We have obtained expressions for the total energy and
the release energy which are valid in 3D as well as in effective
1D and 2D, and the release energy was shown to agree well with
experimentally measured values.  Our results lead to analytic
expressions for the condensate wavefunction and density, as well
as the condensate dimensions for different aspect ratios in
cylindrical geometry. We have also gauged the variation of the
quasi-particle energy spectrum as the condensate deviates from
effective 1D or 2D, which should provide a better estimate of
spectrum-dependent physical quantities in such regimes. Our
variational ansatz should be useful in studying 2D lattices of
effective 1D condensates such as described in \cite{Greiner}; each
of those 1D condensates also has an axial confinement and hence is
exactly the type of high aspect ratio cigar-shaped condensates we
consider here. A study of such lattices is a topic of some of our
ongoing study.

It is a pleasure to acknowledge the support and help of Tom
Bergeman during the initial part of this work and also valuable
conversations with M. G. Moore, H. T. C. Stoof and E. M. Wright.
Part of this work was done with support from ONR, and from NSF
through a grant for the Institute for Theoretical Atomic and
Molecular Physics at the Harvard-Smithsonian Center for
Astrophysics. A later stage of the work was completed with support
from ONR grant N00014-99-1-0806.

\end{document}